\documentclass{emulateapj}

\shorttitle{A Subaru/Suprime-Cam Survey of M31's spheroid along the South-East minor axis}
\shortauthors{Tanaka et al.}

\begin{document}


\title{A Subaru/Suprime-Cam Survey of M31's spheroid along the South-East minor axis\altaffilmark{1}}


\author{Mikito~Tanaka\altaffilmark{2,3},
        Masashi~Chiba\altaffilmark{4},
        Yutaka~Komiyama\altaffilmark{2},
        Masanori~Iye\altaffilmark{2},
        and
        Puragra~Guhathakurta\altaffilmark{5}}


\altaffiltext{1}{Based on data collected at the Subaru Telescope, which is
 operated by the National Astronomical Observatory of Japan.}
\altaffiltext{2}{National Astronomical Observatory of Japan, 2-21-1 Osawa,
 Mitaka, Tokyo 181-8588, Japan; miki@optik.mtk.nao.ac.jp}
\altaffiltext{3}{University of Tokyo, 7-3-1 Hongo, Bunkyo-ku, Tokyo 113-0033,
 Japan}
\altaffiltext{4}{Astronomical Institute, Tohoku University, Aoba-ku, Sendai
 980-8578, Japan}
\altaffiltext{5}{University of California Observatories/Lick Observatory,
 University of California
Santa Cruz, 1156 High Street, Santa Cruz, California 95064, USA}


\begin{abstract}
We have used Suprime-Cam on the Subaru Telescope to conduct a $V$- and
$I$-band imaging survey of fields sampling the spheroid of the Andromeda
galaxy along its south-east minor axis.  Our photometric data are deep enough
to resolve stars down to the red clump. Based on a large and reliable sample
of red giant stars available from this deep wide-field imager, we have
derived metallicity distributions vs.\ radius and a surface brightness profile
over projected distances of $R=23$--66~kpc from the galaxy's center.
The metallicity distributions across this region shows a clear high mean
metallicity and a broad distribution ([Fe/H]$ \sim -0.6 \pm 0.5$), and
indicates no metallicity gradient within our observed range. The surface
brightness profile at $R>40$~kpc is found to be flatter than
previously thought.
It is conceivable that this part of the halo
samples as yet unidentified, metal-rich substructure.
\end{abstract}



\section{Introduction}

Our understanding of how disk galaxies like our own or the Andromeda galaxy
(M31) formed plays a key role in near-field cosmology \citep{Freeman2002},
because such nearby galaxies offer us detailed views of galactic structures
through their resolved stars. In particular, the components of the extended
low surface brightness stellar halo, old field stars and globular clusters,
provide invaluable information on early chemo-dynamical evolution of disk
galaxies over the past $\gtrsim10$~Gyr (i.e., well before thin disk
components appeared).  The spatial distribution of galactic halos suggests
that proto-galaxies were much larger than the sizes of currently bright
galactic disks, possibly a result of hierarchical assembly of subgalactic
systems orbiting at larger radii \citep{SZ1978}.
The importance of accretion and merging of small
systems in the galaxy formation process is also evident in the spatial and
kinematic substructures in the Galaxy's stellar halo
\citep[e.g.,][]{Yanny2003,Helmi1999}.

The volume density profile of the Milky Way halo is characterized by an
$r^{-3.5}$ power-law of Galactocentric radius, based on
direct counts of halo tracers or their orbital motions
\citep[e.g.,][]{Harris1976}.
By contrast, Andromeda's spheroid appears to have a complex structure.
In their imaging study \citet{Pritchet1994} found that the surface
brightness (SB) profile along the minor axis is characterized by a
de~Vaucouleurs $R^{1/4}$ law over projected distances $R$ of
$1\arcsec$ to $1\fdg5$ (few pc to 20~kpc) from the M31 center.
Later, \citet{Guhathakurta2005} reported, based on spectroscopic selection
of red giant branch (RGB) candidates in several halo fields, that
the SB beyond $R\simeq20$~kpc shows a flatter profile with $R^{-2.3}$,
and that the halo may extend to $R\gtrsim150$~kpc.
\citet{Irwin2005} also found, from their photometry of RGB stars over
$R=20$--55~kpc along the minor axis, a comparably
flat SB profile that is fit by an $R^{-2.3}$ power law or an exponential law
with a scalelength of 14~kpc.
It remains to be seen whether this flat portion of
the halo profile is affected by substructure, and if it is dominated
by either metal-rich or metal-poor populations. The latter issue is especially
important since spectroscopic studies suggest that M31's inner spheroid/halo
has a radial gradient in mean metallicity, whereby metal-poor stars
dominate at $R\gtrsim50$~kpc \citep[e.g.,][]{Kalirai2006}. Detailed studies
based on large numbers of stars (say from wide-field imagers) are
required to investigate the fundamental nature of M31's halo along
its minor axis.

In this Letter, we report on our Subaru/Suprime-Cam wide-field photometric
observations of M31's halo aimed at obtaining statistically significant
SB and metallicity distribution (MD) profiles along the south-east (SE)
minor axis of the galaxy. Our imaging survey is optimized to extract halo
profiles in a region that is well outside
the bright disk component and beyond the inner spheroid with
the $R^{1/4}$ brightness law.

\section{Observations and Data Reductions}\label{sec:obs}
\begin{figure}
\includegraphics[width=8.5cm,angle=0]{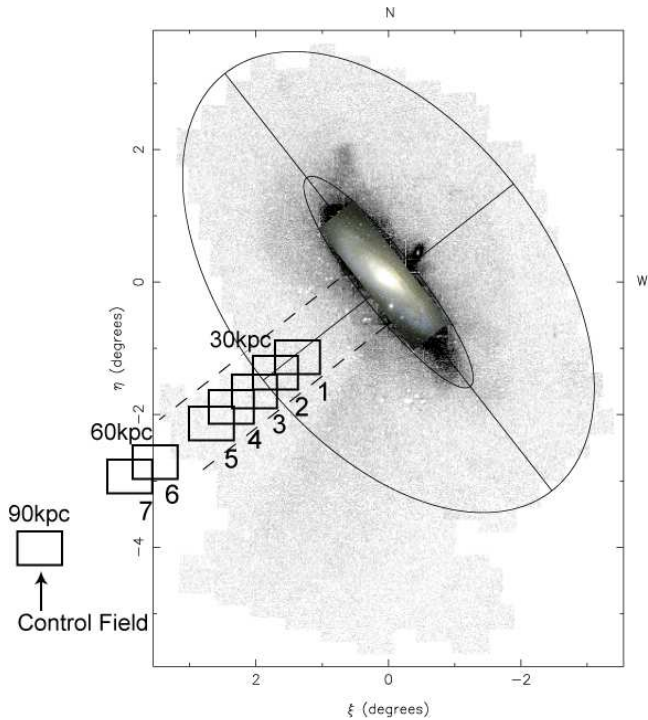}
\caption{
Locations of our Subaru/Suprime-Cam fields (rectangular areas), overlaid
on the surface density map of M31's red RGB stars from the INT/WFC survey of
\citet{Ferguson2002} and adapted from Figure~1 of \citet{Irwin2005}.
Adjacent Suprime-Cam fields overlap by about 25~\% to ensure photometric
calibration.
In addition, the fields used by \citet{Irwin2005} for their surface
brightness profile are delineated by the dashed lines.}
\label{map}
\end{figure}

Using the Suprime-Cam \citep{Miyazaki2002} imager on the Subaru Telescope 
\citep{Iye2004}, which consists of ten $2000\times4000$ CCDs with a
resolution of $0\farcs202$ per pixel and covers a total field-of-view
$34\arcmin \times 27\arcmin$, we have carried out a wide-field imaging survey
of M31's spheroid. Our targeted fields are located between 23 and 
66~kpc from the M31 center (Fig.~\ref{map}).

During four nights in August 2004, we obtained images of seven spheroid
fields (hereafter referred to as Fields~1--7) and a control field in 
Johnson $V$ and Cousins $I$ bands with 
typical seeing FWHM of $0\farcs9-1\farcs3$. 
These observations were obtained in non-photometric conditions.
We carried out additional imaging of the same fields on a photometric night
in August 2005 in order to calibrate the data.
The data were reduced with the software package SDFRED, a useful
pipeline developed to optimally deal with Suprime-Cam images
\citep{Yagi2002,Ouchi2004}. 
Photometry of our co-added images was calibrated through \citet{Landolt1992}
standards, correcting for an airmass term and a color term in each field. 

We then conducted PSF-fitting photometry using the IRAF version of
the DAOPHOT-I\hspace{-.1em}I software \citep{Stetson1987}.
We adopted a 3~$\sigma$ detection threshold for the initial object
detection/photometry pass and iterated the PSF-fitting photometry twice with
5~$\sigma$ and 7~$\sigma$ detection thresholds, respectively, in order to
account for blended stars.
A stellar PSF template was derived from about 100 bright, isolated stars per
image.  Finally, we merged two independent $V$- and $I$-band catalogs into
a combined catalog using a 1-pixel matching radius. 

It is worth noting that the morphological segregation method cannot
distinguish between M31 halo stars and and other point sources such as
Galactic foreground dwarf stars, and compact extragalactic objects.
To statistically remove these contaminations from the targeted fields,
we adopted a control field located at a Galactic latitude comparable to that 
of our targeted fields, on the assumption that it has the same abundance
of foreground and background objects as that in the M31 spheroid fields.

Reddening corrections are applied in each field 
based on the extinction maps of \citet*{Schlegel1998}, 
and the \citet{Dean1978} reddening law $E(V-I)=1.34E(B-V)$ and
$A_I=1.31E(V-I)$. We adopt the M31 Cepheid distance
modulus of $(m-M)_0 = 24.43 \pm 0.06$ ($\sim$ 770 kpc) from
\citet{Madore1995}.
We also note that our analysis
of the SB and MD profiles below is unaffected by photometric
incompleteness: we target RGB stars brighter than $I \sim$ 22.5 (mag), while
our data are more than 90\% complete down to $I \sim$ 23 (mag).  As an
example, even if we count stars down to the 50\% completeness for the $V$
band, the hitherto missing faintest and reddest stars are included in the
analysis but the SB and MD profiles are changed by less than a few percent. 

\section{Results}\label{sec:results}
\subsection{Metallicity Distributions}\label{sec:mdf}

Taking advantage of the large stellar samples available with the wide field
of view of Suprime-Cam, we derive MDs of RGB stars in M31's spheroid, 
by comparing them to RGB fiducials defined by
Galactic globular clusters in the same manner as \citet{Bellazzini2003}. 
For a reliable secure determination of MDs using these templates, we select
RGB stars with
$-3.8 < M_I < -2.0$ and $0.9 < (V-I)_0 < 4.0$. 
These selection criteria allow us to remove a number of contaminants,
such as M31 asymptotic giant branch (AGB) stars, AGB bump stars, and young
stars, along with foreground and background objects.
We perform an interpolation procedure to obtain metallicities of
objects in the spheroid and control fields, and subtract the derived
MD of the control field from that of the spheroid 
field in order to statistically remove the effects of the foreground and
background contaminants.
For more detailed description, see \citet{Tanaka2007}.

\begin{figure}
\includegraphics[width=8.5cm,angle=0]{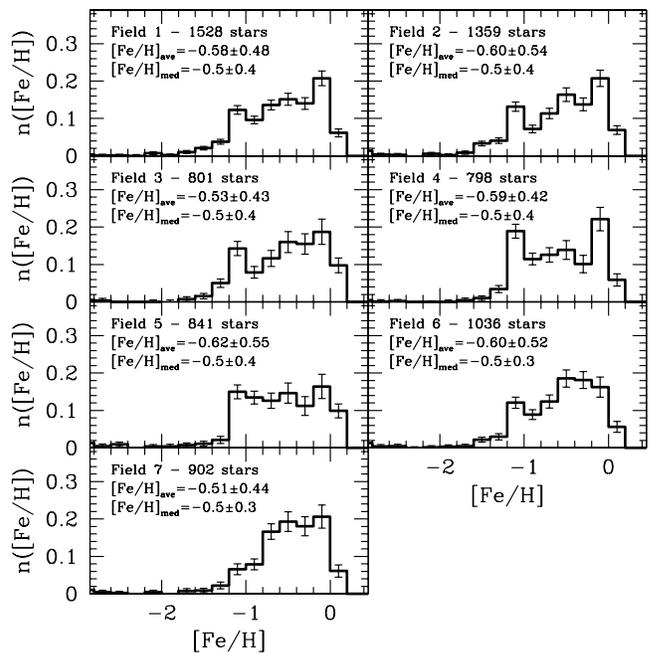}
\caption{Histograms of the metallicity distributions, in the \citet{CG97} scale, 
for the SE minor axis fields of M31's spheroid. 
In the upper left corner of each panel, we show the name of the field, 
the number of stars used to derive the MD, the average metallicity 
([Fe/H]$_{\rm ave}$) together with the associated standard deviation, and
the median metallicity together with the associated semiinterquartile interval.
}
\label{fig:mdf}
\end{figure}

Figure~\ref{fig:mdf} shows the MDs in Fields 1 to 7 of M31's spheroid. 
The vertical error bars denote a nominal uncertainty
in each metallicity bin, yielding from the Poisson errors equal to
$\pm \sqrt{N({\rm spheroid} + {\rm control})}$. 
It is worth noting that because of a large number of the RGB stars
available from our Suprime-Cam data,
the MDs obtained here have significantly small errors and thus are expected
to be most likely compared with the results of the previous studies, which
examined more inner parts of the spheroid than ours
\citep{Durrell2001,Bellazzini2003,Durrell2004}.

It follows that the MDs have a broad distribution ranging from 
[Fe/H] $\sim -3$ to the near solar metallicity. 
The average metallicity ([Fe/H]$_{\rm ave}$) with a standard deviation  
and median metallicity ([Fe/H]$_{\rm med}$) 
with a quartile deviation are shown in each panel. 
In addition, our Field 2 includes the field observed by \citet{Durrell2004}, 
and the MD of Field 2 is almost consistent with their MD.

\begin{figure}
\includegraphics[width=8.5cm,angle=0]{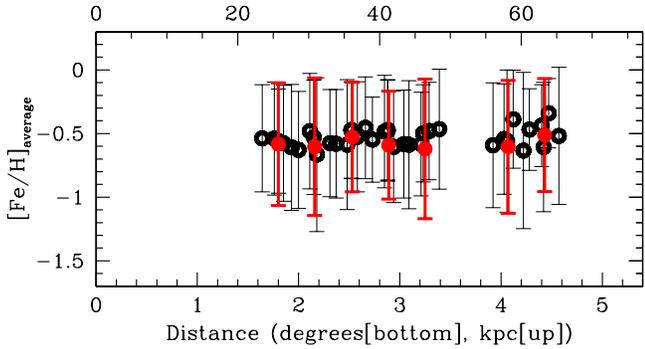}
\caption{The average metallicity plotted as a function of distance from
the M31 center. 
Filled red circles present the average values of metallicity in the seven
targeted fields, while open circles denote those in each subfield
inside a Suprime-Cam field (see text).
}
\label{fig:mdg}
\end{figure}

The most striking feature of the MDs shown in Fig.~\ref{fig:mdf} 
is their overall similarity. For the sake of comparison of them, 
we show their average metallicities as a function of distance 
from the M31 center in Figure~\ref{fig:mdg}. Seven filled red circles 
present the average values of metallicity in our targeted fields,
while many open circles are those in the subfields, which are obtained by
dividing a single Suprime-Cam field into six fields 
(4000 pixels $\times$ 3000 pixels per divided field). 
These subfields help us to examine the fine spatial variation
in a Suprime-Cam field. Conversely, a filled red circle reflects the mean value
of metallicities of six subfields.
Vertical error bars show standard deviation of the mean. 
These plot suggests that the stellar content of the spheroid 
has no metallicity gradient within our observational range. 
This may be consistent with the idea of chaotic merging of small building blocks
first proposed by \citet{SZ1978} for understanding the lack of a metallicity
gradient in the globular cluster system of the Milky Way halo.

\subsection{Surface Brightness Profiles}\label{sec:sb}

We estimate the SB profiles of M31's spheroid along the SE minor axis using
the RGB stars, which are brighter than the AGB bump ($I \sim$ 23.3 mag). After extracting
the secure RGB stars on which are imposed the same magnitude
and color criteria as in preceding section, 
we divide them into the following two groups using the RGB templates 
of Galactic globular clusters \citep{Bellazzini2003}: 
{\it Metal-Poor group} (MP) defined as $-2.16<$[Fe/H]$<-1.11$
and {\it Metal-Rich group} (MR) defined as
$-1.11<$[Fe/H]$<0.07$.
The values of $-2.16$ and $0.07$ are both ends of the templates, and 
the template having a value of $-1.11$ is chosen because it nearly 
marks a discontinuity in the MDs obtained here. 

In contrast to the inner dense parts of the halo in M31, it is possible to
resolve the individual stars in its outer parts, because of their sparse
density and of the sufficiently close distance of M31 to us.
Thus the SB profiles in spheroid's outer part are available by directly
counting individual stars. The sources of noise in this method of estimating
the SBs arise from the Poisson statistics from the finite number of stars observed
and contamination from Galactic stars along the line of sight to the spheroid.
The contribution from unrelated Galactic stars can be estimated by observing
other nearby fields: we use our control field at $\sim$ 90 kpc to this end.

\begin{figure}
\includegraphics[width=8.5cm,angle=0]{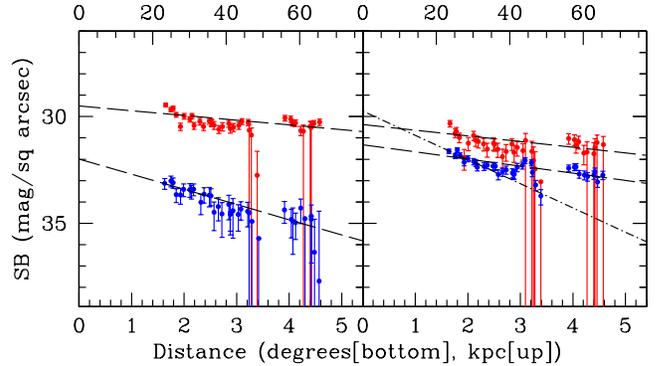}
\caption{The left panel shows the $V$- and $I$-band minor-axis profiles
for RGB stars as derived from the MP and MR group, while the right panel shows
those based on the selection method of RGB stars by \citet{Irwin2005}.
The $V$-band and $I$-band profiles are illustrated with blue and red points,
respectively, which are derived from star counts in the magnitude and
color selection boxes as described in the text.
The error bars reflect a combination of Poissonian and background uncertainties.
The dashed lines show an exponential profile with a scalelength of $s=21.9$ kpc
for MP and $s=70.0$ kpc for MR (left panel), 
and $s=46.4$ kpc for blue RGB stars and $s=58.5$ kpc for red RGB stars (right panel).
The dot-dashed line in the right panel shows an exponential profile 
with $s = 13.7$ kpc by \citet{Irwin2005}, for the sake of comparison.
}
\label{fig:sb}
\end{figure}

The left panel of Figure~\ref{fig:sb} shows the SB profiles 
of the MP (blue) and MR (red) group, which are used to derive the $V$- and 
$I$-band profiles, respectively. The dashed lines show an exponential SB profile,
$\exp( - R / s )$, with a scale length $s$ of 21.9 kpc for MP and 70.0 kpc for MR.
We note that at large distances $R$, the number of RGB stars is just slightly larger
than that of background objects, so we can determine only the upper limit for
the most of their errors. It is remarkable that the SB in the MP group decreases
more steeply with $R$ than that in the MR group, suggesting
that as above stated there may exist a metal-rich substructure at $R = 50 \sim 60$ kpc. 

The right panel of Fig.~\ref{fig:sb} shows the SB profiles based on the
different star selection by \citet{Irwin2005}: using $V$-band and Gunn $i$-band
systems, they selected blue RGB stars with $20.5<i<22.5$ and $24.85 - 2.85(V-i) <
i < 26.85 - 2.85(V-i)$ and red RGB stars with $21<i<22$ and $i > 26.85 -2.85(V-i)$,
and derived the $V$- and $I$-band profiles from these categorized stars,
respectively. We shift $I$-band magnitude criterion about 0.02~mag toward
the fainter side in order to match the Gunn $i$-band system.
We note that our selection criteria for MP and MR are more advantageous
than this fixed magnitude and color selection of blue and red RGB stars
for investigating the effects of the MDs on the SBs.
To compare with the blue RGB star count profile in \citet{Irwin2005},
which is fitted to an exponential profile with $s = 13.7$ kpc, we shift 
it vertically to roughly overlap with our blue RGB surface brightness profile 
in the radial range $1\fdg6-3\fdg0$. This result shows that stellar density 
at around 45 kpc and 60 kpc significantly exceeds a mean halo profile
obtained by \citet{Irwin2005}. Hence, it is possible that
two as-yet-unknown substructures consist of bright metal-rich stars. 

We note that the details of the SB profiles obtained here are found to be pretty
insensitive even if we adopt much fainter $V$- and $I$-band magnitude limits,
say down to $V = 25$ mag and $I = 24$ mag. Thus, the current results are
real and robust against the effects of any contaminations over the fields;
our deep Suprime-Cam data enable us to assess the minor effects of these errors.

\section{Discussion and Conclusions}\label{sec:summary}

As fossil records of galaxy formation, M31's spheroid offers us a global
perspective about its spatial structures as well as stellar populations.
Based on a wide-field imager of Subaru, we have confirmed the previous
results \citep{Guhathakurta2005,Irwin2005} that the SB profile in the SE minor
axis is much flatter at $R > 23$ kpc than the innermost part of the spheroid
described by the $R^{1/4}$ law. Our finding of no metallicity gradient in this
outer part over $R = 23 \sim 66$ kpc implies that dissipationless processes,
such as accretion and merging of collisionless stellar systems, may have been
at work for the formation of this spheroidal part, thereby erasing any metallicity
nonuniformity. We note here that our control field at $R = 90$ kpc of the SE
minor axis, which shows no substructures \citep{Ibata2007}, 
may still contain some spheroidal stars, because M31's spheroid may
extend up to $R \sim 150$ kpc \citep{Guhathakurta2005}. However, even so, the
relative MD and SB profiles obtained here remain unaltered. 

Intriguingly, our finding of the possible overdensity regions at around 45 and
60 kpc is in agreement with the recent work by
\citet{Ibata2007} (having appeared in astro-ph while
preparing for the submission of our paper), who also found the stream-like
features at similar locations, referred to as stream D and C, respectively.
Our metallicity analysis suggests that their stream D appears to be slightly
more metal-poor than stream C. Also, the MDs of stream C (our Field 6 and 7)
have a high-metallicity peak similar to those of
the Giant Stream \citep{Tanaka2007}. However, it is yet uncertain if stream C
is indeed related to the Giant Stream because of the lack of the information
on the distance to stream C as well as its internal kinematics.
Deeper photometry down to horizontal-branch magnitude and/or multi-object
spectroscopy of stars will be important for clarifying the origin of
the newly detected substructures. 

Our results reported here are not totally inconsistent with the finding
by \citet{Kalirai2006} and \citet{Chapman2006}, who argue for the presence
of metallicity gradient obtained from their kinematical studies of the bright 
RGB stars using Keck/DEIMOS. Indeed, only our Field 2 overlaps Field a0 of
\citet{Kalirai2006}, where both fields yield basically similar metallicity, and
other fields of Field 2 to 7 along the minor axis do not overlap their survey
regions (see their Figures 1 and 12). It is interesting to remark that their
Field m6 at $R \simeq 87$ kpc along the minor axis shows rather high
metallicity of [Fe/H]$\sim -0.85$, being similar to that of their Field a0
as well as ours. In contrast, other fields deviated from the minor axis (such as
their a13, a19, and b15) show low metallicities, which basically
give rise to the reported metallicity gradient. Thus, this implies that
the MDs along the minor axis fields in concern may be somewhat different from
other outer halo fields, i.e., more higher surface density and more metal-rich
than previously thought as reported here. A tempting view for these
properties of our observed fields is that there exists a metal-rich
substructure at $R$ of 50-60 kpc, corresponding to a faint tail of
the giant stellar stream \citep{Ferguson2002} or a part of another stream.
More extensive studies by observing large halo areas of Andromeda, e.g., using
much wider HyperSuprime camera under construction, will be necessary
to arrive at more decisive conclusions and thus obtain the accurate
formation picture of this typical disk galaxy.

\acknowledgments

We thank the Subaru Telescope staff for the excellent support during
our observing runs. Data reduction/analysis was carried out on ``sb'' computer
system operated by the Astronomical Data Analysis Center of the National
Astronomical Observatory of Japan.

\end{document}